\documentstyle
[12pt]{article} \hoffset -0.5in \textwidth 6.5in \textheight
9.00in  \parskip 7pt \openup2.5\jot \parindent=0.5in
\newskip\humongous \humongous=0pt plus 1000pt minus 1000pt
\def\caja{\mathsurround=0pt}
\def\eqalign#1{\,\vcenter{\openup1\jot \caja
        \ialign{\strut \hfil$\displaystyle{##}$&$
        \displaystyle{{}##}$\hfil\crcr#1\crcr}}\,}
\newif\ifdtup

\def\eqright #1\cr{\noalign{\hfill$\displaystyle{{}#1}$}}
\def\eqleft #1\cr{\noalign{\noindent$\displaystyle{{}#1}$\hfill}}

\def\oldreffmt#1{\rlap{[#1]} \hbox to 2\parindent{}}

\def\figfmt#1{\rlap{Figure {#1}} \hbox to 1in{}}

\def\holdtheequation{\arabic}
\def\sectioneq{\def\holdtheequation{\thesection.\arabic}\let
\section={\section\setcounter{equation}{0}}\setcounter
{equation}{0}\def\theequation{\holdtheequation{equation}}}
\newcounter{holdequation}

\def\begineq #1\endeq{$$ \refstepcounter{equation}\eqalign{#1}\eqno
        (\theequation) $$}
\def\contlimit{\,{\hbox{$\longrightarrow$}\kern-1.8em\lower1ex
\hbox{${\scriptstyle (a\rightarrow0)}$}}\,}
\def\centeron#1#2{{\setbox0=\hbox{#1}\setbox1=\hbox{#2}\ifdim
\wd1>\wd0\kern.5\wd1\kern-.5\wd0\fi
\copy0\kern-.5\wd0\kern-.5\wd1\copy1\ifdim\wd0>\wd1
\kern.5\wd0\kern-.5\wd1\fi}}
\def\centerover#1#2{\centeron{#1}{\setbox0=\hbox{#1}\setbox
1=\hbox{#2}\raise\ht0\hbox{\raise\dp1\hbox{\copy1}}}}
\def\centerunder#1#2{\centeron{#1}{\setbox0=\hbox{#1}\setbox
1=\hbox{#2}\lower\dp0\hbox{\lower\ht1\hbox{\copy1}}}}
\def\lsim{\;\centeron{\raise.35ex\hbox{$<$}}{\lower.65ex\hbox
{$\sim$}}\;}
\def\gsim{\;\centeron{\raise.35ex\hbox{$>$}}{\lower.65ex\hbox
{$\sim$}}\;}

\def\super#1{\ifmmode \hbox{\textsuper{#1}}\else\textsuper{#1}\fi}
\def\textsuper#1{\newcount\holdspacefactor\holdspacefactor=\spacefactor
$^{#1}$\spacefactor=\holdspacefactor}

\def\supercite{\def\cite{\newcite}}

\def\newcite#1{\super{\newcount\citenumber\citenumber=0\getcite#1,@, }}
\def\getcite#1,{\advance\citenumber by1
\def\getcitearg{#1}\def\lastarg{@}
\ifnum\citenumber=1
\ref{#1}\let\next=\getcite\else\ifx\getcitearg\lastarg\let\next=\relax
\else ,\ref{#1}\let\next=\getcite\fi\fi\next}

\def\pom{{\rm P\kern -0.53em\llap I\,}}
\def\spom{{\rm P\kern -0.36em\llap \small I\,}}
\def\sspom{{\rm P\kern -0.33em\llap \footnotesize I\,}}

\relax

\def\super#1{\ifmmode \hbox{\textsuper{#1}}\else\textsuper{#1}\fi}
\def\textsuper#1{\newcount\holdspacefactor\holdspacefactor=\spacefactor
$^{#1}$\spacefactor=\holdspacefactor}

\def\supercite{\def\cite{\newcite}}

\def\newcite#1{\super{\newcount\citenumber\citenumber=0\getcite#1,@, }}
\def\getcite#1,{\advance\citenumber by1
\ifnum\citenumber=1
\ref{#1}\let\next=\getcite\else\ifx#1@\let\next=\relax
\else ,\ref{#1}\let\next=\getcite\fi\fi\next}

\supercite
\def\bibitem{\item\label}%

\renewcommand{\thefootnote}{\fnsymbol{footnote}}

\def\mainhead#1{\setcounter{equation}{0}\addtocounter{section}{1}
  \vbox{\begin{center}\large\bf #1\end{center}}\nobreak\par}

\supercite

\begin{document} \begin{titlepage}
\rightline{\vbox{\halign{&#\hfil\cr
&ANL-HEP-CP-92-117\cr
&BROWN-HET-885\cr
&\today\cr}}}
\vspace{0.25in}
\begin{center}

{\Large\bf
THE $\eta_6$ AND MASSIVE PHOTON PAIRS AT LEP
}\footnote{Work supported by the U.S. Department of
Energy, Division of High Energy Physics, Contract\newline W-31-109-ENG-38.}
$$
\vspace{.1in}
$$
\normalsize  Alan R. White
\footnote{Presented at the American Physical Society, DPF 92, Fermilab,
November 10-14, 1992.} and Ian G. Knowles
 \\
\smallskip
High Energy Physics Division\\Argonne National
Laboratory\\Argonne, IL 60439\\
\medskip
\medskip
\normalsize  Kyungsik Kang\\
\smallskip
Physics Department\\Brown University\\Providence RI 02912\\\end{center}

\begin{abstract}
The $\eta_6$, a ``heavy axion''  associated with sextet quark electroweak
symmetry breaking, may have been seen at LEP via its two-photon decay mode and
also at TRISTAN via its hadronic decay modes.

\end{abstract}

\renewcommand{\thefootnote}{\arabic{footnote}} \end{titlepage}

\mainhead{1. INTRODUCTION}

Dynamical electroweak symmetry breaking by a chiral condensate of
{\it color sextet quarks}\cite{sex} has many attractive features,
including the following.

\begin{itemize}

\item[{i)}] Color plays the role of technicolor - the electroweak scale is
a $QCD$ scale.

\item[{ii)}] The symmetry breaking is that of a ``minimal'' Higgs sector
- ensuring ``$\rho = 1$''.

\item[{iii)}] There is a resolution of the Strong $CP$ problem via a
unique ``heavy axion'' - the $\eta_6$ - that should be looked for
instead of the Higgs.

\item[{iv)}] The necessary sextet quark sector may play a vital role
within $QCD$ giving, in particular, unitary Critical Pomeron scaling at
high energy.

\end{itemize}
This last feature underlies the persistent advocacy of sextet symmetry
breaking by one of us (ARW) for more than ten years. Nevertheless our
understanding of the subtle dynamics of the sextet sector remains poor.
Some existing, but unexplained, phenomena that may be ``hints'' of its
existence include:
\begin{itemize}
\item[{1)}] the strong production of $W^+W^-$ and $Z^0Z^0$ pairs - UA1 ($W$
+ 2 jet) events,
\item[{2)}] composite operator contributions in the CDF inclusive jet
cross-section,
\item[{3)}] the anomalous behaviour of the $\bar{p}p$ real part - as perhaps
seen by UA4,
\item[{4)}] ``exotic'' Cosmic Ray events which suggest a new strong interaction
scale.
\end{itemize}
About 3 years ago, we noted\cite{ka} that the Geminion and mini-Centauro exotic
events could be explained as the diffractive production of a heavy particle
and suggested this could be the $\eta_6$ (the estimated mass is now\cite{ari}
$\geq$ 50 GeV). We linked the corresponding threshold to the UA4 result
and proposed looking for the $\eta_6$, {\it via its two photon decay
mode}, in hadronic diffractive interactions. Subsequently it was pointed
out\cite{hu} that $Z^0 \to \eta_6 + \gamma$ could be observed (as a rare
decay) at LEP.

\mainhead{2. IMPLICATIONS OF LEP AND TRISTAN RESULTS}

As reported at this meeting L3, DELPHI and perhaps ALEPH, have seen
several events of the form $Z^0 \to l^+l^- + \gamma\gamma$ in each of
which the mass of the $\gamma\gamma$ pair is close to 60 GeV. The
lepton pairs are either muons or electrons. There are, as yet, no
neutrino, quark, or $\tau$ pairs. While the kinematics of some of
the events may be compatible with $QED$ radiation, others look
implausible explained this way. Therefore, the events suggest the
existence of a massive ``particle'' with a mass of, perhaps, 59 GeV and a
width of O(1) GeV. The first presumption would surely be that the new
particle is radiated directly by the $Z^0$ with the lepton pair coming
from an off-shell $Z^{0*}$ (although we shall discuss another
possibility later).

It is also very interesting to ask whether this ``particle'' has been
seen at TRISTAN. In fact all three experiments saw a peak at 59.05 GeV.
AMY obtained\cite{amy} a value of R more than $30\%$ above
the standard model value (although with a large error - giving at most a
``2$\sigma$'' effect). If this is produced by the same new particle that
appears in the LEP events, we can infer both that it
couples to $e^+e^-$ and that it {\it has major hadronic decay modes}.

If the width were due to electroweak couplings to quark and lepton
states, in analogy with the $Z^0$, then such decays would surely have
already been seen at LEP. It seems more likely that high multiplicity
hadron states are involved which would not be so easily identified at LEP,
but clearly would be registered at TRISTAN. We emphasise therefore that

{\bf the existence of a massive particle with appreciable couplings to the weak
bosons and non-trivial strong decays (not proportional to quark masses),
strongly suggests electroweak symmetry breaking involves constituents that
carry color.}

Sextet quark symmetry breaking is an obvious candidate and an
immediate question is whether the new particle could be the $\eta_6$?
At first sight, as we discuss, the production rates involved appear to
be calculable from the sextet quark triangle anomaly and are too low by
orders of magnitude. However, as we shall also discuss, the Strong
$CP~(and~C)$ properties of the sextet quark sector are subtle and sextet
quark Goldstone boson amplitudes violate $CP$. As a result there exist
large ``longitudinal'' $Z^0$ and $W^{\pm}$ amplitudes which can not be
calculated from the
anomaly but could give large enough cross-sections at LEP and TRISTAN.

\mainhead{3. COLOR SEXTET SYMMETRY BREAKING AND THE $\eta_6$
AXION}

A massless flavor doublet $(U,D)$ of color sextet quarks with the usual
quark quantum numbers (except that the role of quarks and antiquarks is
interchanged) is first added to the Standard Model {\it with no scalar
Higgs sector}. Within $QCD$, conventional chiral dynamics will
break the sextet axial flavor symmetries spontaneously and produce four
massless pseudoscalar mesons (Goldstone bosons), which we denote as
$\pi^+_6, \pi^-_6, \pi^0_6$ and $\eta_6$. The $\pi^+_6, \pi^-_6$, and $\pi^0_6$
are ``eaten'' by the massless electroweak gauge bosons and respectively become
the third components of the massive $W^+$, $W^-$ and $Z^0$, giving $M_W
\sim g~F_{\pi_6}$ where $F_{\pi_6}$ is {\it a $QCD$ scale}.
$F_{\pi_6} \sim 250 GeV$ is consistent with an elementary ``Casimir Scaling''
rule\cite{sex}.

The $\eta_6$ is not involved in generating mass for the electroweak
gauge bosons. Instead it is a ``Peccei-Quinn axion'' which produces
Strong $CP$ conservation in the triplet quark sector. Its mass is much
higher than conventionally expected for an axion because of the intricate
$QCD$ dynamics of the sextet sector. The evolution of $\alpha_s$ is
negligible above the electroweak (sextet) scale and there is an
effective infra-red fixed-point controlling the dynamics. The absence of
the infra-red growth of the gauge coupling implies that, in the sextet
sector, confinement and chiral symmetry breaking involve instantons as
an important ``infra-red'' effect. (There are no renormalons and
instantons don't melt!). Each instanton interaction contains a factor\cite{hol}
of $\cos~[\tilde{\theta}~+<\eta_6>]$, where (in a conventional notation)
$\tilde{\theta}~=~\theta~+~\det~m_3$, and an axion potential of the form
$V(\cos~[\tilde{\theta}~+~<\eta_6>])$ is generated. Such a potential naturally
retains the $CP$-conserving minimum at $~\tilde{\theta}~+~<\eta_6>~=~0$
while also giving an $\eta_6$ mass (the curvature at the minimum) of
order the electroweak scale - say 60 GeV!

\mainhead{4. ANOMALY AMPLITUDES AND LONGITUDINAL COUPLINGS}

If we assume that $CP$ is conserved also in the sextet sector
we obtain a set of amplitudes which are far too small to be
compatible with the LEP events. In analogy with the chiral dynamics of the
physical pion sector, $PCAC$ allows us to calculate\cite{hu} vertices
for $\eta_6 \to \gamma\gamma$, $Z^0 \to \eta_6 + \gamma$ and $Z^0 \to
\eta_6 + Z^{0*}$ directly from the sextet quark triangle anomaly. For
$\eta_6 \to \gamma\gamma$ we obtain a very narrow width of 0.17 keV. (If
the full experimental width is much larger then this would imply that a
large fraction of the hadronic cross-section at LEP must involve the
$\eta_6$!) For $Z^0 \to \eta_6 + \gamma$ we would predict one event in 20
million at LEP. While from $Z^0 \to \eta_6 + Z^{0*}$ we obtain a rate
for $Z^0 \to \eta_6 + \mu^+\mu^-$ of 2 events in a billion. This is at
least three orders of magnitude too small to explain the two photon events.

Note, however, that the $\eta_6$ acts as an axion {\it only in
the low energy effective lagrangian for the triplet quark sector}. The
full $QCD$ lagrangian for the combined triplet and sextet sectors has no
axion. Also to generate triplet quark (and lepton) masses, four-fermion
($CP$-violating) couplings must be added. As a consequence the ``low-energy''
effective lagrangian for $QCD$ interactions of the $\eta_6$, $\pi^+_6$,
$\pi^-_6$, and $\pi^0_6$ is {\it necessarily $CP$-violating}. In
unitary gauge, it is the ``longitudinal'' (or scalar) components
of the gauge boson fields i.e. $\partial^{\mu}Z^0_{\mu}$,
$\partial^{\mu}W^+_{\mu}$ and $\partial^{\mu}W^-_{\mu}$, that inherit the
interactions of the Goldstone bosons $\pi^0_6$, $\pi^+_6$ and $\pi^-_6$
respectively. Therefore such interactions give large, $CP$-violating, couplings
of the form $\eta_6\partial^{\mu}Z^0_{\mu}\partial^{\mu}Z^0_{\mu}$,
$\eta_6\partial^{\mu}W^+_{\mu}\partial^{\mu}W^-_{\mu}$,
$\eta_6\partial^{\mu}Z^0_{\mu}\partial^{\mu}W^+_{\mu}\partial^{\mu}W^-_{\mu}$
.. etc..

\mainhead{5. $\eta_6$ DECAY MODES}

To compute electroweak amplitudes for the $\eta_6$ a first step would be
to write an effective lagrangian for the (unitary gauge) longitudinal
amplitudes we have just described, add the electroweak interaction, and
compute to the lowest order in the electroweak couplings. This gives some
immediate order of magnitude estimates.

First we note that $\eta_6 \to \gamma\gamma$ is given by a
$\partial^{\mu}W_{\mu}$ loop which, because of the unitary gauge
propagators, is dominated by large momenta. $\eta_6 \to l^+l^-$ is given
by a similar loop but with one $\partial^{\mu}W_{\mu}$ propagator
replaced by a neutrino propagator. Both amplitudes are O($\alpha_{ew}$) but
the photon amplitude should be significantly larger because of the additional
unitary gauge boson propagator. The $\eta_6 \to e^+e^-$ vertex can
potentially be large enough to allow the $\eta_6$ to be seen at TRISTAN.

Photon emission is generally favored by its coupling directly, to the Goldstone
bosons involved, at large momentum. The $e^+e^- \to e^+e^- +
\gamma\gamma\gamma$
event recorded by L3, in which the $\gamma\gamma\gamma$ mass is also close to
60 GeV., could involve an $\eta_6 \to \gamma\gamma\gamma$ decay.

Consider now the hadronic decay modes of the $\eta_6$.
Perturbative gluon emission exposes the large sextet quark constituent
mass ($\sim 400 GeV$) and so is very suppressed. We expect instanton
interactions to be responsible for the dominant hadronic decays. The simplest
possibility would be an isotropic distribution of five quarks and
five antiquarks (one of each flavor), giving a high multiplicity hadron state
with many (mini-) jets. (There is some, statistically insignificant, evidence
in the published data on multiplicity distributions\cite{amy} that the
increased cross-section at Tristan is of this form.) At LEP, the combination
of such a state with a hard lepton pair (i.e. $m_{l^+l^-} \sim 20-30 GeV$), as
in the two photon events, should be quite distinctive.

\mainhead{6. LEPTON PAIRS IN THE LEP EVENTS}

While it seems that the two-photon component of the LEP events can be
straightforwardly associated with the decay of the $\eta_6$, the lepton
pairs are not so simple. If the events are indeed produced by
$Z^0 \to \eta_6 + Z^{0*} \to [\gamma\gamma] + [l^+l^-]$ then, because $CP$ is
not conserved, the $Z^{0*}$ can be longitudinal and the initial vertex can be
very large. However, the lepton vertex would then contain a factor $m_l$ ,
giving {\it no neutrino pairs}, a negligible number of electron
pairs, and an overwhelming number of $\tau$ pairs, compared to muon
pairs! Since quark pairs carry color, $QCD$ interactions with the
initial vertex could reduce the corresponding amplitude for
their production. Nevertheless, it seems unlikely that a simple mass
dependence of the lepton vertex will correctly describe the experimental
situation as more events appear.

There is, however, a further source of lepton pairs. The four-fermion
sextet/lepton couplings, that provide masses in combination with the
sextet condensate, will also provide a direct (short-distance) coupling
of lepton pairs into (small) instanton interactions - without going via the
electroweak interaction. That is {\it massive leptons} can be
directly produced out of the sextet quark interaction by a mechanism
that is {\it closely related to, but not identical to}, the
mass-generation mechanism. Giving, in particular, $Z^0 \to \eta_6 +
l^+l^-$ vertices. This mechanism will also produce lepton pairs out of
very hard $QCD$ interactions in hadron colliders.

\mainhead{7. CONCLUSIONS}

The $\eta_6$ is a strong candidate for the new particle suggested both by the
massive photon pairs seen at LEP and by TRISTAN data at the
corresponding energy.

Unambiguous ``discovery'' of the $\eta_6$ could be a fundamental
breakthrough in the problems of electroweak symmetry breaking and Strong
$CP$ conservation - at current accelerators!

It (and other signals of the sextet sector) should also be looked for in hadron
colliders, in diffractive interactions and very high energy hard collisions.

\end{document}